# Frequency-resolved Raman Thermometry Analysis via a Multi-layer Heat Transfer Model for Bulk and Low-dimensional Materials

Taocheng Yu[1], Yilu Fu[1], Chenguang Fu[2], Tiejun Zhu[2], Wee-Liat Ong,[1,3]*

[1]ZJU-UIUC Institute, College of Energy Engineering, Zhejiang University, Haining, Jiaxing, Zhejiang 314400, China

[2]State Key Laboratory of Silicon and Advanced Semiconductor Materials, School of Materials Science and Engineering, Zhejiang University, 310058 Hangzhou, China.

[3]State Key Laboratory of Clean Energy Utilization, Zhejiang University, Hangzhou Zhejiang, 310027, China

## Abstract

Raman thermometry is advantageous for measuring the thermal transport of low-dimensional materials due to its non-contact nature. Transient Raman methods have improved the accuracy of steady-state Raman thermometry by removing the need for accurate temperature calibration and laser absorption evaluation. However, current methods often resort to finite element analysis (FEA) to decipher the measured signals. This step is time-consuming and impedes its ubiquitous adaptation. In this work, we replace the FEA by fitting the transient-state Raman signal to a three-dimensional (3D) analytical heat transfer model for measuring the thermal conductivity of two bulk layered materials [i.e., molybdenum disulfide ($MoS_2$) and bismuth selenide ($Bi_2Se_3$) crystals] and the interfacial thermal conductance ($h$) of CVD-grown $MoS_2$ and molybdenum di-selenide ($MoSe_2$) on quartz ($SiO_2$). Our measured results agree reasonably well with literature and theoretical calculations. We also performed a quantitative sensitivity analysis to give insights on how to improve the measurement sensitivity. Our work provides an efficient way to process the data of transient-based Raman thermometry for high throughput measurements.

**Keywords:** Raman, thermal conductance, thermal conductivity, 2D, interface, layered

## 1. Introduction

Characterizing the thermal properties of two-dimensional (2D) materials is an ongoing challenge due to their atomic thickness, rendering them sensitive to environmental influences [1–3]. The optothermal Raman technique is advantageous for this task due to its non-contact nature [4]. In this technique, temperature-dependent Raman properties, including peak position, linewidth, and a ratio of Stokes and anti-Stokes intensity, can be used as a thermometer to probe a local temperature through a steady-state or transient measurement. The steady-state method [5] often requires accurate calibrations between the temperature and the Raman signal (i.e., the temperature coefficient of the Raman signal), with precise knowledge of the amount of absorbed laser power. These requirements are often the sources of errors for this method [6]. To overcome these drawbacks, several transient methods using pulse or square-wave lasers were developed recently, including the time domain differential Raman (TD-Raman) [7], energy transport state-resolved Raman (ET-Raman) [8], frequency-resolved Raman (FR-Raman) [9,10] and frequency-domain energy transport state-resolved Raman (FET-Raman) [6]. In these methods, a short laser irradiation induces a transient temperature rise, changing the resultant Raman signal. This quasi-steady-state signal is normalized and compared to theoretical predictions calculated using a brute-force scan of different parameters, to obtain the best values for the targeted parameters. Finite element analysis (FEA) is often used to provide theoretical predictions that simulate the transient temperature responses in these methods [6,11,12]. However, the need for numerous different simulations and the long simulation time, especially in high-frequency heating cases for reaching a quasi-steady-state, makes the data processing and sensitivity analysis difficult, hindering the popularity of these Raman techniques. While analytical models for Raman thermometry exist, they are often tailored to 1D heat transfer in specific geometries [7,9] or the 3D process in the steady-state laser-flash Raman method [13–15] thus lacking generalizability.

The frequency domain thermo-reflectance (FDTR) technique is another powerful thermometry method for the thermal transport measurements of nanoscale materials [16–18]. It often employs a continuous-wave laser with a sinusoidally-modulated intensity to excite a periodic temperature signal on a sample while using an unmodulated laser to probe the resulting thermo-reflectance signal. One of its main advantages is the fast and easy determination of the unknown thermal properties in a multilayered material system by fitting the signal to an analytical heat transfer equation. Its main disadvantages for measuring the thermal properties of 2D materials include the need for a smooth metal transducer coating that will affect the intrinsic thermal conductivity of the atomically thick films and the difficulty of separating the thermal properties of 2D materials from its interfacial thermal conductance with the substrate or transducer [19]. Transducer-less FDTR is an alternative implementation, but the weak absorbance and reflectance of the laser by the measured sample may result in a weak and low-quality thermo-reflectance signal [20]. Hence, the thermal properties of 2D materials are predominantly measured using Raman methods [21].

In this paper, we modified the 3D analytical heat transfer model from the FDTR method to process the data from the FR-Raman technique for a multilayered system. Our model decomposes the square wave heating profile used in the FR-Raman into a series of sinusoidal waves and recomposes the resulting temperature responses to fit the normalized measured Raman signal for extracting the unknown thermal properties. Using this model, we successfully measured and analyzed the in-plane thermal conductivity of two bulk layered-materials, $MoS_2$ and $Bi_2Se_3$, and the interfacial thermal conductance of two 2D materials, $MoS_2$ and molybdenum di-selenide ($MoSe_2$) supported on $SiO_2$.

## 2. Materials and methods

### 2.1. Materials

The $MoS_2$ single crystal was purchased from SixCarbon. The $Bi_2Se_3$ bulk polycrystal was prepared as follows. Highly pure element chunks of Bi, Se (5N, Emei Semiconductor Materials Research Institute) were weighted according to the stoichiometric ratio of $Bi_2Se_3$. The mixtures were sealed into clean quartz tubes below 10-3 Pa and melted in the Muffle furnace at 1273 K for 12 h. The tubes were rocked twice to ensure composition homogeneity during melting and finally cooled in air to obtain ingots, which were then mechanically milled into powders by planetary ball milling at 400 rpm for 3 h in the Ar atmosphere. The powders were sintered into dense bulks by spark plasma sintering under 723 K and 80 MPa for 30 min. Before the measurement, we exfoliated away the surface layers using scotch tape to expose fresh sample surfaces. The CVD-grown $MoS_2/SiO_2$ and $MoSe_2/SiO_2$ were purchased from Nanjing

MKNANO Tech. Co. Ltd. and Beike 2D Materials Co. Ltd., respectively. Each sample has 10±2 layers and an area coverage more than 96%.

## 2.2. Frequency-resolved Raman Thermometry

Our method employed an electro-optic modulator (EOM) to modulate the laser intensity into a square wave (SW) with "ON" and "OFF" states. The "ON" state corresponds to a constant-intensity laser heating that induces a temperature rise on the irradiated sample. The average temperature rise is measurable from the Raman spectra. During the "OFF" state, as no laser irradiates the sample, the sample naturally cools down, resulting in no Raman spectrum. A normalized temperature rise $\Phi$ can be calculated using Equation (1):

$$\Phi = \frac{\Theta_2}{\Theta_1} = \frac{T_2 - T_0}{T_1 - T_0} = \frac{(\partial T / \partial \omega) \cdot (\omega_2 - \omega_0)}{(\partial T / \partial \omega) \cdot (\omega_1 - \omega_0)} \tag{1}$$

where $T_0$ is the initial (room) temperature, $T_1$ and $T_2$ are Raman-measured average temperature at two modulation frequencies ($f_1$ and $f_2$), $\Theta_1$ and $\Theta_2$ are the corresponding temperature rises at these two frequencies, $\omega_0$ is the Raman shift at room temperature, $\omega_1$ and $\omega_2$ are the Raman shifts under SW laser heating. Performing this normalization procedure minimizes the effect of laser absorbance and temperature coefficient on the final results [9].

## 2.3. Experiemental details

The Raman spectra were collected using a confocal Raman microscopy (Renishaw inVia Basis) system with an excitation laser of wavelength 532 nm and a 20X objective lens (NA = 0.4). The laser spot radius is measured to be 4μm using a knife-edge method (see Figure S1). The room temperature Raman shift $\omega_0$ in Eq. (1) is determined by varying the power of a CW laser to heat the sample to obtain the corresponding Raman spectra. A red shift occurs as the laser power increases (see Figure S2). The Raman shifts are linearly related to the laser power $P$ for the two peaks of our samples. We can determine $\omega_0$ by linearly fitting the $P$-$\omega$ relationship to get the intercept at zero laser power. The laser power is carefully selected to be large enough to suppress the noise while small enough to avoid damaging the samples.

## 3. Results

### 3.1. Heat transfer model

The FEA simulation is usually performed to relate $\Phi$ and $f$ [10]. After running numerous time-consuming numerical simulations to a quasi-steady-state with different values for the thermal properties of interest, the simulated result that best matches the experiment signal can be found. Hence, a long simulation time may be needed, especially at high modulation frequencies. Here, we modified the heat transfer model used in the FDTR technique to replace these numerical simulations. The FDTR technique often utilizes a sinusoidally intensity-modulated pump beam to heat the sample, inducing a sinusoidal temperature response. Here, we decompose the square wave heating profile into a series of equivalent sine waves using the Fourier transform and input them into the heat transfer model. By summing up the resulting temperature signals, we can obtain the corresponding periodic temperature response induced by this square wave heating.

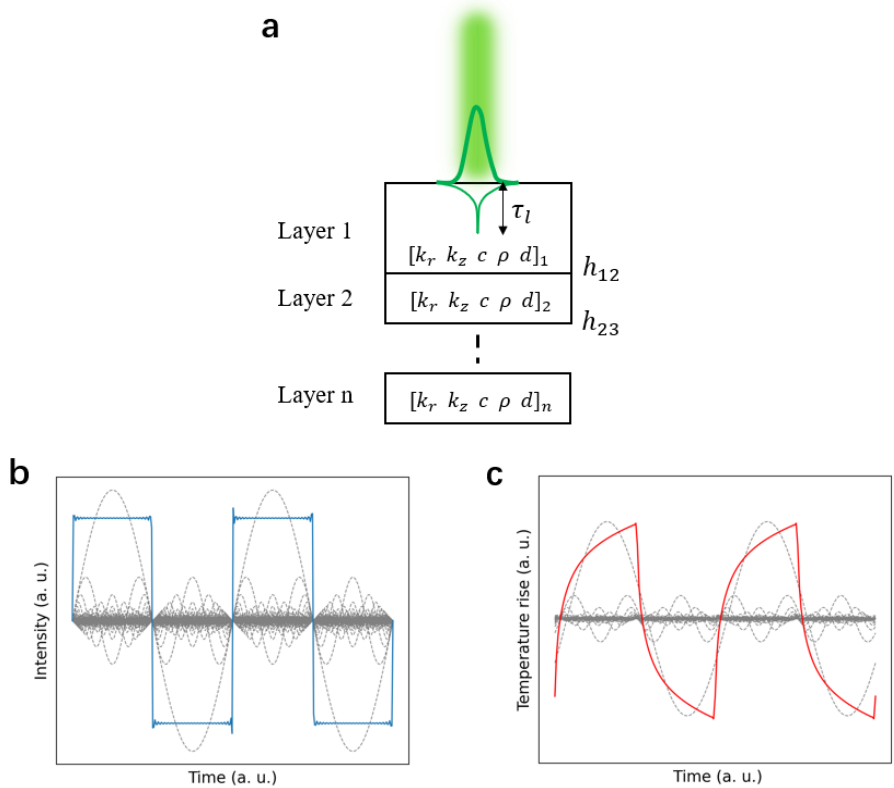

Figure 1. (a) Schematics of multilayered sample heated by a Gaussian laser beam with its optical penetration depth. Thermophysical properties for each layer are shown, including in-(out-of-) plane thermal conductivity $k_r$ ($k_z$), specific heat $c$, mass density $\rho$, thickness $d$,

interfacial thermal conductance $h$, and optical penetration depth $\tau_l$. Schematics of (b) decomposition of a square wave (blue solid line) into sinusoidal waves (grey dashed lines) and (c) re-composition of the temperature responses of sinusoidal waves (grey dashed lines) into that of a square wave (red solid line). There are 100 sinusoidal waves in (b) and (c).

For a multilayered system shown in Figure 1(a), we assume the laser is only absorbed in the first layer. The laser intensity profile is described as a Gaussian beam in the radial direction that decays exponentially into the sample, as shown below,

$$\dot{g}(r,z,t) = \frac{2P(t)}{\pi r_0^2 \tau_l [1-\exp{(-\frac{d_1}{\tau_l})}]} \exp{(-\frac{2r^2}{r_0^2})} \exp{(-\frac{z}{\tau_l})} \tag{2}$$

where $P$ is the laser power absorbed by the first layer, $r_0$ is the $1/e^2$ spot radius, $\tau_l$ is the optical penetration depth, and $d_1$ is the thickness of the first layer. The governing heat equation in the first layer can be written as,

$$\frac{k_r}{r}\frac{\partial}{\partial r}\left(r\frac{\partial T}{\partial r}\right) + k_z\frac{\partial^2 T}{\partial z^2} + \dot{g}(r,z,t) = C_p\frac{\partial T}{\partial t} \tag{3}$$

where $k_r$ and $k_z$ are the in-plane and out-of-plane thermal conductivity and $C_p$ is the volumetric heat capacity.

We decompose our periodic square wave heating input into a series of sinusoidal wave inputs using a discrete Fourier transform [22],

$$f(x) = \sum_{n=0}^{N/2} a_n \cos\left(\frac{2\pi n x}{T'}\right) + b_n \sin\left(\frac{2\pi n x}{T'}\right) \qquad x = 0,1,2\dots,N \tag{4}$$

$$\begin{cases} a_n = \frac{2}{T'}\int_0^{T'} f(x)\cos\left(\frac{2\pi n x}{T'}\right)dx \\ b_n = \frac{2}{T'}\int_0^{T'} f(x)\sin\left(\frac{2\pi n x}{T'}\right)dx \end{cases} \tag{5}$$

where $T'$ is the period of a square wave and $N$ is the number of sampling points in one period. Each of the resulting sinusoidal waves applied to the above heat transfer model will return a sinusoidal temperature response with amplitude $A_n$ and phase $\varphi_n$, similar to the FDTR solution for anisotropic materials [23]. Finally, the temperature response of the square wave is formed by superposing these sinusoidal temperature responses,

$$T(x) = \sum_{n=0}^{\frac{N}{2}} a_n * A_n * \cos\left(\frac{2\pi n x}{T} + \varphi_n\right) + b_k * A_n * \sin\left(\frac{2\pi n x}{T} + \varphi_n\right), \ x = 0,1,2\dots,N \tag{6}$$

The schematics of the decomposition of a square wave and the recomposition of temperature responses is shown in Figure 1(b) and (b), respectively.

*3.2. Measurement results for bulk layered-materials*

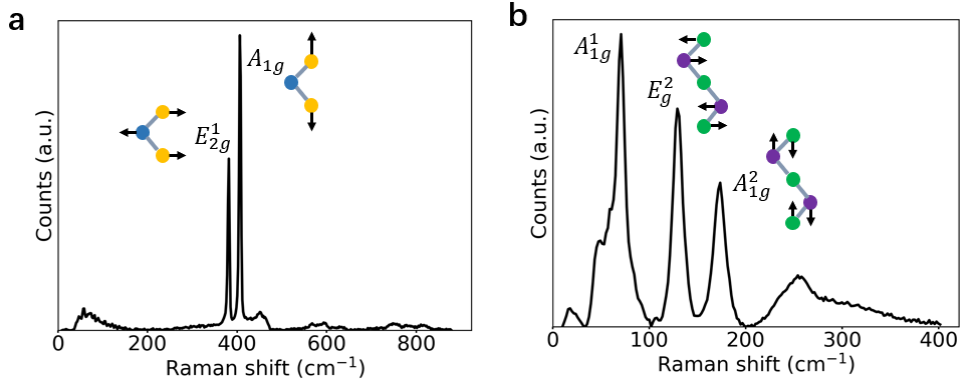

Figure 2. Our measured Raman spectrum of bulk (a) MoS₂ and (b) Bi₂Se₃. The insets show the corresponding vibrational modes.

We first validated the solutions of our analytical model with FEA simulations. The excellent agreement with the FEA results gives us the confidence to use Eq. (6) to analyze the signal from a transient Raman thermometry experiment (Figure S3). We first measured the in-plane thermal conductivity of two prototypical bulk layered-materials, MoS₂ and Bi₂Se₃. The Raman spectra of the two samples are shown in Figure 2(a) and (b). Two sharp peaks, i. e. $E^1_{2g}$ and $A_{1g}$, can be seen in the MoS₂ Raman spectrum, representing the in-plane and out-of-plane vibrational modes, respectively. For Bi₂Se₃, three relatively sharp peaks can be seen, namely, $A^1_{1g}$, $E^2_g$ and $A^2_{1g}$. Although the $A_{1g}$ has the highest intensity, it overlaps with other peaks. Therefore, we select the $E^2_g$ and $A^2_{1g}$ modes for subsequent measurements.

Data is taken using logarithmically spaced frequency points from kHz to approximately 8 MHz. Typical fitting results are shown in Figure 3, together with the normalized temperature rise $\Phi$ as black dots. The data points are fitted to our heat transfer model using a least square method with the in-plane thermal conductivity $k_r$ as the only free parameter. The optical penetration depth $\tau_l$ in the analytical model is calculated from $\tau_l = \frac{\lambda}{4\pi k_L}$, where $\lambda$ is the laser wavelength (532 nm) and $k_L$ is the extinction coefficient of the material [6]. The input parameters and the initial values for $k_r$ are listed in Table 1. From Table 2, the fitted $k_r$ of both MoS₂ and Bi₂Se₃ agree with the literature values regardless of the Raman peaks used. The error bars of Bi₂Se₃ are longer than that of MoS₂ as the former can only withstand a smaller amount of irradiation before reaction, resulting in a lower signal-to-noise ratio. Regardless, the $\pm 20\%$ uncertainty curves can be distinguished from the best-fit curve, which suggests a sufficiently large sensitivity of $\Phi$ to $k_r$.

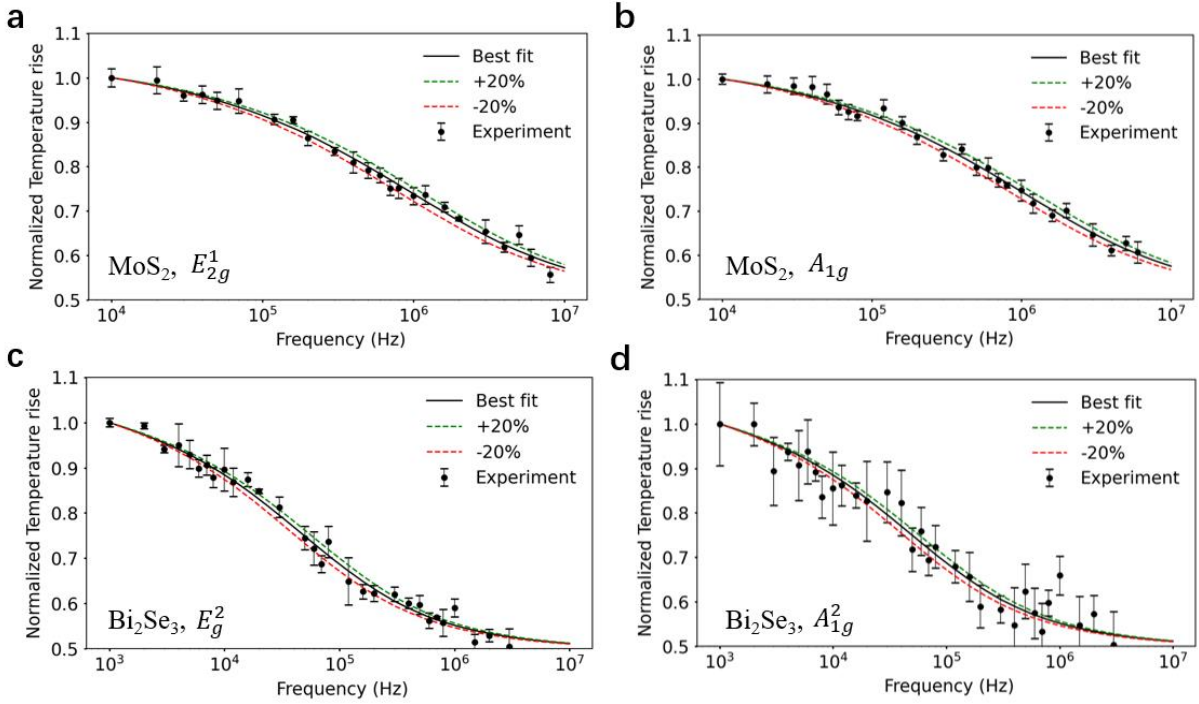

Figure 3. Fitting results for in-plane thermal conductivity of bulk MoS$_2$ (a) with $E_{2g}^1$ mode (b) with $A_{1g}$ mode, and in-plane thermal conductivity of bulk Bi$_2$Se$_3$ (c) with $E_g^2$ mode (d) with $A_{1g}^2$ mode. The error bar on each point is calculated from five consecutive measurements. The best fits are shown as the black lines, while the green and red dashed lines represent the $\pm 20\%$ uncertainty from the best fit.

**Table 1** Parameters used for data fitting and sensitivity analysis.

| Parameters | $k_r$ | $k_z$ | $\tau_l$ | $c$ | $r_0$ |
|---|---|---|---|---|---|
| Units | W/(mK) | W/(mK) | nm | J/(kgK) | $\mu m$ |
| MoS$_2$ | 80.0 [24] | 4.75 [24] | 14 [25] | 382 [24] | 4.0 |
| Bi$_2$Se$_3$ | 3.1 [26] | 0.7 [27] | 19 [28] | 205 [27] | 4.0 |
| Uncertainty | NA | 20% | 20% | 5% | 5% |

**Table 2** Measured in-plane thermal conductivity and their corresponding literature values.

| Samples | Mode | Our work [W/(mK)] | Literature [W/(mK)] |
|---------|------|-------------------|---------------------|
| MoS$_2$ | $E_{2g}^1$ | $86.7 \pm 11.6$ | 80 [24] |
|  | $A_{1g}$ | $85.7 \pm 12.7$ |  |
| Bi$_2$Se$_3$ | $E_g^2$ | $3.05 \pm 0.34$ | 3.1 [26] |
|  | $A_{1g}^2$ | $2.84 \pm 0.48$ |  |

*3.3. Sensitivity analysis for bulk materials*

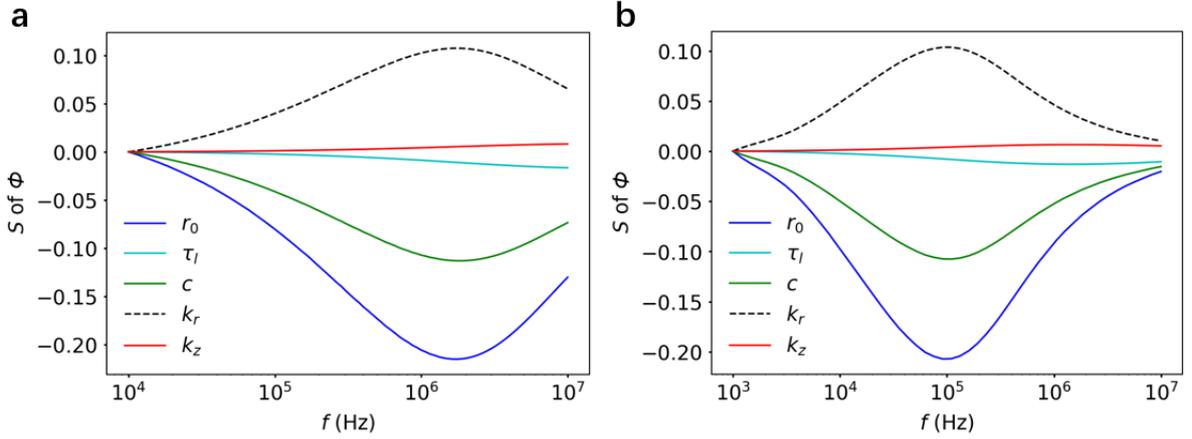

Figure 4. Sensitivity of $\Phi$ to the thermal properties of bulk (a) MoS$_2$ and (b) Bi$_2$Se$_3$

To better understand the sensitivity ($S$) of $\Phi$ to the parameters in the model, we calculated

$$S = \frac{\partial \ln \Phi}{\partial \ln x} \tag{13}$$

where $x$ is the parameter of interest. We analyze the $S$ to the following parameters: spot radius ($r_0$), optical penetration depth ($\tau_l$), specific heat capacity ($c$), and in-plane/out-of-plane thermal conductivity ($k_r$, $k_z$), with their nominal values listed in Table 1. The sensitivity results of bulk MoS$_2$ and Bi$_2$Se$_3$ in Figure 4(a) and (b) depict that the out-of-plane thermal conductivity $k_z$ has a near-zero sensitivity throughout the frequency range, allowing us to assume a value. The optical penetration depth $\tau_l$ also has low sensitivity and is thus fixed (Table 1). The low sensitivity to $k_z$ and $\tau_l$ comes from the sizable difference between the optical and thermal penetration depth. The thermal penetration depth ($d_{pz}$) of a sinusoidally-intensity-modulated laser [20] can be approximated as $d_{pz} = \sqrt{\frac{k_z}{\pi c \rho f}}$. Since the amplitude of the square wave is dominated by the sine wave of the same frequency [see Figure 1(c)], we assumed the $d_{pz}$ of this sine wave for the square wave, which is 280 nm and 126 nm at 10 MHz for MoS$_2$ and Bi$_2$Se$_3$. As the $d_{pz}$ is much larger than $\tau_l$ (i.e., 10~20 nm), the temperature gradient in the depth direction is relatively gentler when compared to the laser intensity gradient near the sample surface. The temperature near the surface dominates $\Phi$, as it is weighted by the laser intensity in the depth direction,. Therefore, small changes in $\tau_l$ or $k_z$ (which changes $d_{pz}$) hardly affect the measured signal.

The $k_r$ for both materials have a relatively large sensitivity, producing distinguishable $\pm$ 20% variations from the best-fit curves in Figure 3. The frequency at which $\Phi$ is most sensitive to $k_r$ is 1.84 MHz and 0.1 MHz for MoS$_2$ and Bi$_2$Se$_3$ (Figure 4), with the spot radius $r_0$ having a high sensitivity, as in most laser-based techniques [16,29]. So, an accurate determination for $r_0$ is crucial for measuring $k_r$. Additionally, the specific heat capacity $c$ has a sensitivity comparable to $k_r$. Similar to the analysis for $k_z$ and $\tau_l$, the relative length scale between the laser spot radius and the in-plane heat spreading distance (i.e., $d_{pr} = \sqrt{\frac{k_r}{\pi c \rho f}}$) can explain the correlation between the sensitivity of $k_r$, $r_0$, and $c$. At 1.84 MHz and 0.1 MHz, the $d_{pr}$ of MoS$_2$ and Bi$_2$Se$_3$ are 2.69 $\mu m$ and 2.66 $\mu m$ and comparable to $r_0$. The highest $S$ occurs at $d_{pr}/r_0 \sim 0.67$ for both materials. Variations in $d_{pr}/r_0$ can arise from $k_r$, $r_0$, or $c$, which mirrors the sensitivity correlation among these three parameters. The relationship between $S$ and $d_{pr}/r_0$ also necessitates an appropriate choice of $r_0$ so that frequency with the highest $S$ is included for more accurate measurements (See Supplementary Materials).

### 3.4. Measurement results for 2D materials

Next, we used our method to measure the interfacial thermal conductance $h$ of two multilayered 2D materials on SiO$_2$ substrates (i.e., MoS$_2$/SiO$_2$ and MoSe$_2$/SiO$_2$). The $A_{1g}$ mode, with its higher Raman signal, was chosen over the the $E_{2g}^1$ mode in our measurements. As the transmittance of SiO$_2$ is over 90% [30], we assumed that heat was absorbed only in the 2D materials. Representative fitting results for MoS$_2$/SiO$_2$ and MoSe$_2$/SiO$_2$ are shown in Figure 5(a) and (b), with the interfacial thermal conductance ($h$) as the only free parameter and the values of other relevant parameters listed in Table 3. The $h$ of MoS$_2$/SiO$_2$ and MoSe$_2$/SiO$_2$ are initially measured to be 6.2±1.9 MW/(m$^2$K) and 2.8±0.5 MW/(m$^2$K), respectively. As the optical-acoustic (OA) phonon nonequilibrium can be important in the Raman measurement of 2D materials [31], we corrected for this using the method developed by Hunter et al. [32] (see Supplementary Materials) by subtracting the OA phonon nonequilibrium term from the $\varPhi$ data points in Figure 5(a) and (b) before refitting. After the correction, the $h$ of MoS$_2$/SiO$_2$ and MoSe$_2$/SiO$_2$ becomes 10.9±3.3 MW/(m$^2$K) and 4.3±0.6 MW/(m$^2$K), respectively. The $h$ of MoS$_2$/SiO$_2$ falls within the range of prior studies [0.44~50 MW/(m$^2$K)] [33–35], while the $h$ of MoSe$_2$/SiO$_2$ is higher than prior studies done on exfoliated samples [0.09~0.13 MW/(m$^2$K)] [36].

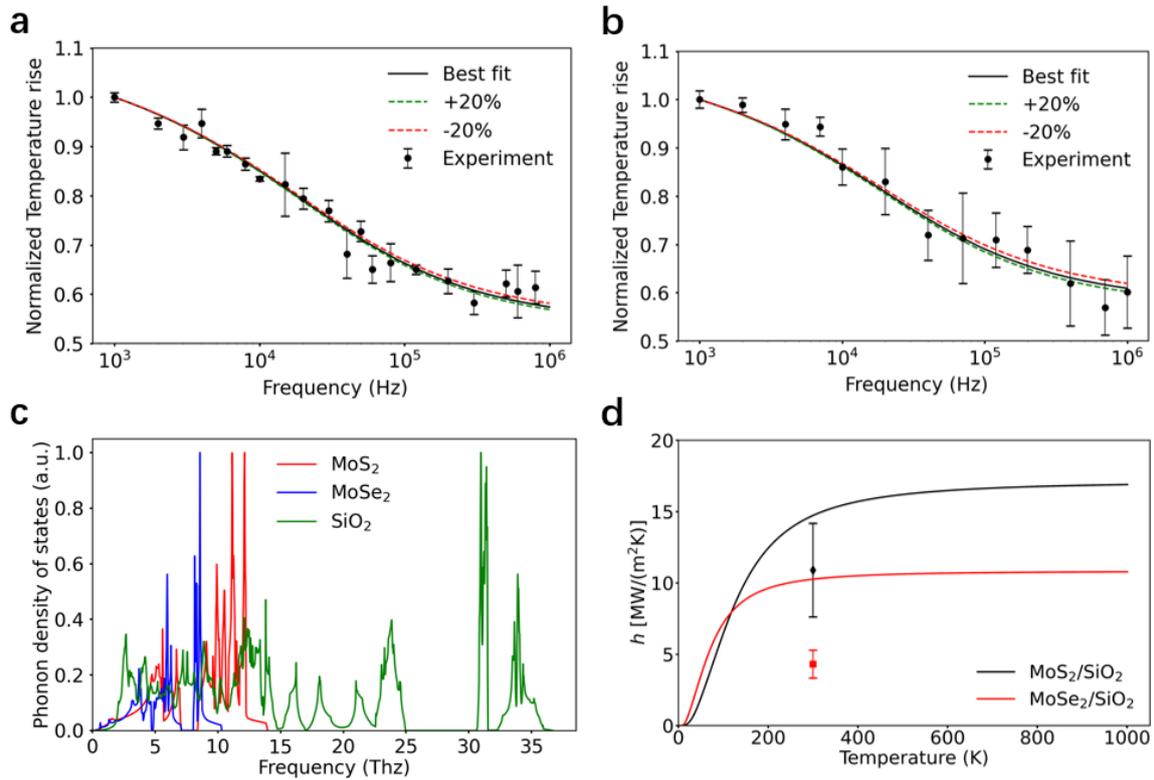

Figure 5. Typical fitting result for (a) MoS$_2$/SiO$_2$ and (b) MoSe$_2$/SiO$_2$. The best fits are shown as the black lines, while the green and red dashed lines represent the ±20% uncertainty of the fitting results. (c) Phonon density of states of MoS$_2$, MoSe$_2$, and SiO$_2$. (d) Interfacial thermal conductance of MoS$_2$/SiO$_2$ and MoSe$_2$/SiO$_2$ from Raman thermometry (squares) and calculated from DMM (solid lines).

To better understand the difference in our measured $h$ trend, we perform theoretical calculations using density functional theory (DFT) to obtain the phonon dispersion and density of states (pDOS) of the three materials involved. As shown in Figure 5(c), the pDOS of $MoS_2/SiO_2$ has a larger overlap area than that of $MoSe_2/SiO_2$, which can promote the transmission of phonons. Using the diffuse mismatch model (DMM) (see details in the Supplementary Materials), our calculated $h$ of $MoS_2/SiO_2$ is also higher than that of $MoSe_2/SiO_2$ at 300K. However, these calculated values are higher than those from our measurements. The over-prediction of $h$ from DMM can originate from several reasons, including an assumed perfect interface and total diffusive phonon scattering [37]. Although there is a gap between our experimental and theoretical results, the measured and theoretical $h$ magnitude and trend at 300K between these two systems are similar, validating our method.

**Table 3** Parameters used for data fitting and sensitivity analysis.

| Parameters | $k_r$ | $k_z$ | $\tau_l$ | $c$ | $r_0$ | $l$ | $h$ | $k_s$ | $c_s$ |
|---|---|---|---|---|---|---|---|---|---|
| Units | W/(mK) | W/(mK) | nm | J/(kgK) | $\mu m$ | nm | MW/($m^2$K) | W/(mK) | J/(kgK) |
| $MoS_2/SiO_2$ | 40 [38] | 4.75 [24] | 14 [25] | 382 [24] | 4.0 | 6 | 10 | 1.36 [39] | 743 [40] |
| $MoSe_2/SiO_2$ | 10 [6] | 3 [6] | 20 [6] | 270 [6] | 4.0 | 6 | 5 | 1.36 [39] | 743 [40] |
| Uncertainty | 20% | 20% | 20% | 5% | 5% | 20% | NA | 5% | 5% |

### 3.5. Sensitivity analysis for 2D materials

The sensitivity curves of the $MoS_2/SiO_2$ and $MoSe_2/SiO_2$ systems are shown in Figure 6. Except for the interfacial thermal conductance, all other properties of the 2D materials have near-zero sensitivities. Thus, $h$ is the only measurable property in such systems. This result can be understood qualitatively using the Biot number ($Bi$) [41]. For the heat transfer across the interface, the $Bi$ can be calculated as $\frac{hl}{k_z}$, where $h$ is the interfacial thermal conductance, $l$ is the characteristic length (thickness of 2D materials), and $k_z$ is the out-of-plane thermal conductivity of 2D materials. The $Bi$ for $MoS_2/SiO_2$ and $MoSe_2/SiO_2$ are 0.014 and 0.0086, respectively. As they are much lower than 0.1, the interfacial thermal resistance dominates over the internal resistance of the 2D materials. On the other hand, the substrate thermal conductivity ($k_s$) and heat capacity ($c_s$) have higher sensitivities, as the substrate thermal resistance is relatively large. We estimate the $Bi$ on the side of the substrate ($Bi_s$) as $\frac{h d_{pz,s}}{k_s}$, with the characteristic length being the thermal penetration depth in the substrate ($d_{pz,s}$). At the modulation frequency of 1 MHz, the $Bi_s$ for $MoS_2/SiO_2$ and $MoSe_2/SiO_2$ is 4.2 and 1.6, respectively. These larger-than-one $Bi_s$ values indicate that the thermal resistance of the substrate dominates the thermal transport over the interfacial thermal resistance.

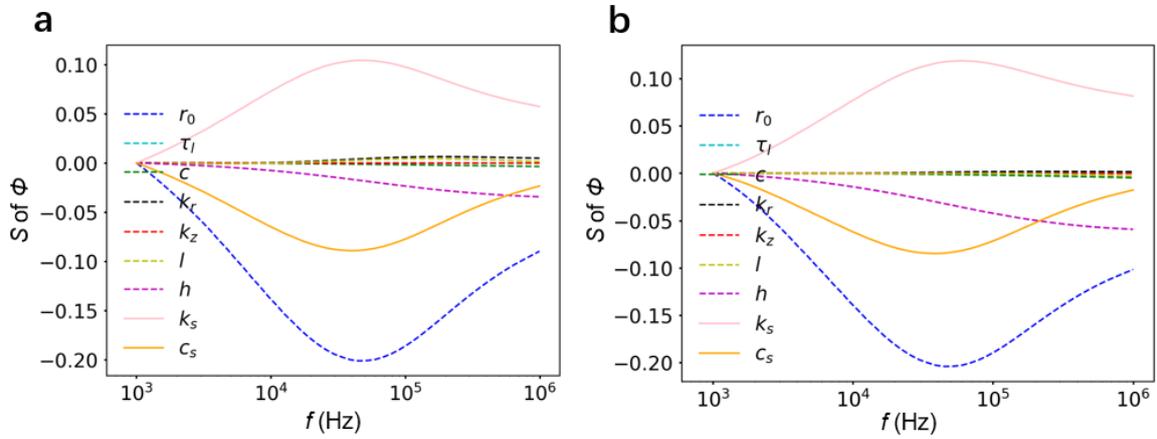

Figure 6. Sensitivity curves for (a) $MoS_2/SiO_2$ and (b) $MoSe_2/SiO_2$. Solid lines are for substrates, while dashed lines are for 2D materials.

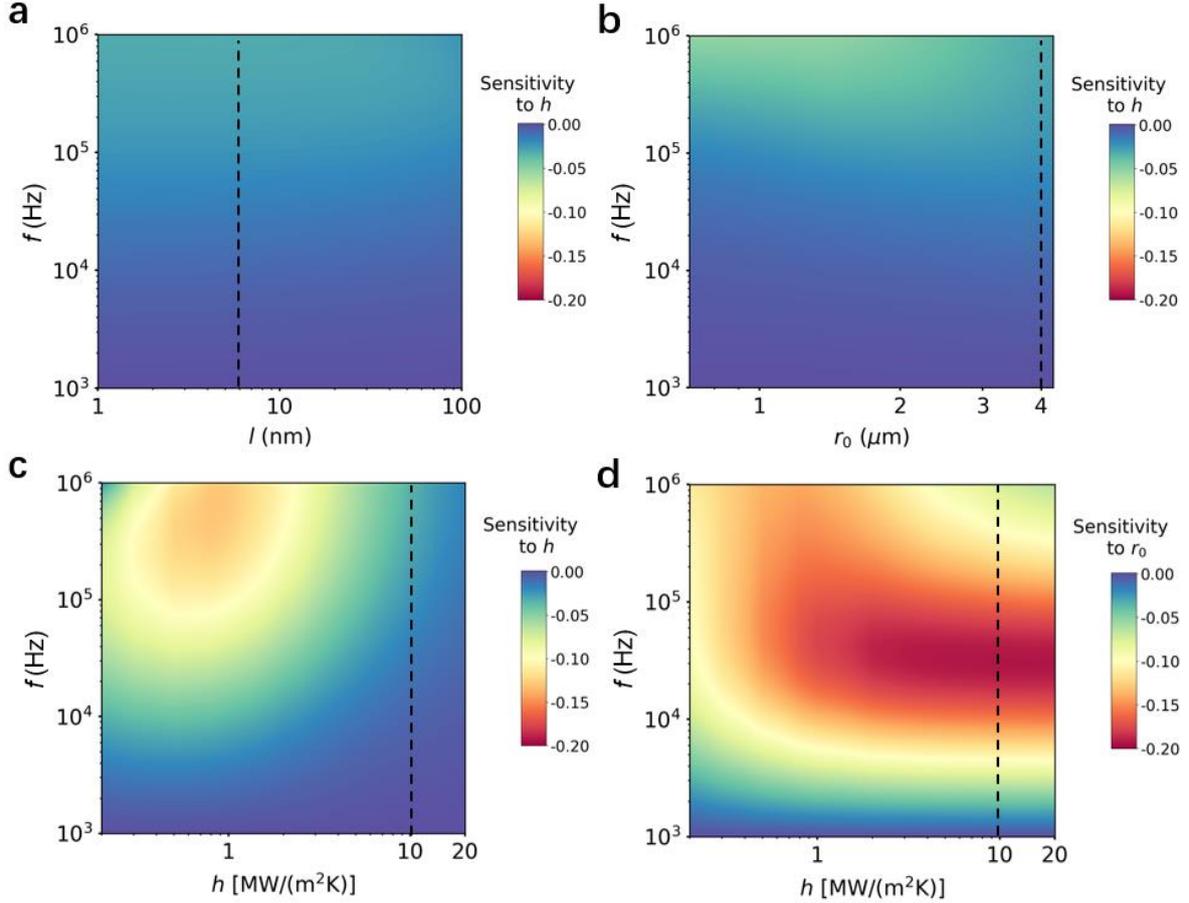

Figure 7. Sensitivity contour of $\Phi$ to $h$ for different (a) sample thickness $l$, (b) spot radius $r_0$, and (c) interfacial thermal conductance $h$. (d) Sensitivity of $\Phi$ to $r_0$ at different interfacial thermal conductance. The black dashed lines represent the sensitivity of our current measured system.

The $h$ is affected by the contact quality, which can vary from sample to sample, depending on the preparation method (CVD/mechanical exfoliation) and the defect concentration (air gaps, wrinkles, etc.) [42]. To understand how the different parameters affect the sensitivity to $h$ ($S_h$), we use the $MoS_2/SiO_2$ as an example. As shown in Figure 7(a), there is no significant change in $S_h$ with thickness (current thickness denoted as the black dashed line). As the internal thermal resistance increases with thicker film, so does the sensitivity to the material properties of the 2D material (Figure S9). As shown in Figure 7(b), a smaller spot radius from a higher magnification objective does not increase $S_h$ significantly, as interfacial thermal transport instead of in-plane thermal transport dominates the temperature profile. The interfacial thermal conductance value itself affects $S_h$. As shown in Figure 7(c), the maximum $S_h$ magnitude increases from 0.04 to 0.13 with a smaller $h$ [0.5 – 1 MW/(m²K)]. On the other hand, the magnitude of the sensitivity to $r_0$ decreases for smaller $h$, as shown in Figure 7(d). This decrease is beneficial for the measurement of $h$ because the uncertainty contributed by $r_0$ will be small.

### 3.6. Limitations of the model

Like other Raman techniques, our method requires the material to have Raman-active modes, temperature-sensitive Raman characteristics, reasonably high light absorbance, chemical stability, etc [21]. Although our heat transfer model enables convenient data processing and sensitivity analysis, it is constructed based on a few important assumptions.

First, the heat source with an exponentially decaying profile is only valid when the hot carrier diffusion length ($L_D$) is small compared to the spot size. The $L_D$ is several hundred nanometers for MoS₂ [43] and Bi₂Se₃ [44], and tens of nanometers for MoSe₂ [6], are much smaller than the laser spot size. So, the effect of hot carrier diffusion in these materials can be neglected [20].

Second, the model assumes local equilibrium among the different thermal carriers as the Raman shift signal only indicates optical phonon temperature. The optical-acoustic phonon non-equilibrium in bulk MoS₂ is insignificant with our large spot radius of 4μm [45,46]. For Bi₂Se₃, its thermal conductivity is contributed by electrons and phonons [47], with strong coupling between these energy carriers [48–50], so the non-equilibrium between these carriers is negligible. On the other hand, the acoustic-optical phonon non-equilibrium can be significant in 2D materials, especially when the spot size is small and needs to be considered.

Third, in our multilayered model, the heat is assumed to be absorbed in the first layer. For 2D materials whose thickness is smaller than the optical penetration depth, we also need to consider the heating effect from the laser energy absorbed by a non-transparent substrate.

## 4. Conclusions

Based on the FR-Raman thermometry, we demonstrated a fast data processing technique, which can also be extended to other transient state Raman methods. The results from a 3D analytical heat transfer model instead of finite element simulations fitted the normalized Raman shift data to extract the unknown thermal properties. The thermal response of a square wave heating profile used in FR-Raman is obtained by superpositioning a series of sinusoidal laser heating responses. After validating our heat transfer model with results from FEA, we measured and reproduced the in-plane thermal conductivity literature values for bulk $MoS_2$ and $Bi_2Se_3$. We furthered our model to measure the interfacial thermal conductance of supported 2D materials systems, $MoS_2/SiO_2$ and $MoSe_2/SiO_2$. The obtained $h$ of $MoS_2/SiO_2$ is higher than that of $MoSe_2/SiO_2$, which agrees with the literature measurements and our theoretical DMM calculations. Sensitivity analyses for the mentioned material systems indicate suitable modulation frequency range and accurate determination of spot radius are crucial for improving measurement accuracy. Also, a smaller interfacial thermal conductance between a 2D material and its substrate provides better accuracy. Our method can, thus, accelerate the measurement and data analysis of the FR-Raman technique and facilitate the high-throughput in-situ measurement/diagnosis for 2D-based micro-devices.

**CRediT authorship contribution statement**

**Taocheng Yu:** Methodology, Investigation, Writing – original draft. **Yilu Fu:** Methodology, Investigation. **Chenguang Fu:** Resources, Writing – review & editing. **Tiejun Zhu:** Resources, Writing – review & editing. **Wee-Liat Ong:** Conceptualization, Supervision, Writing –review & editing, Project administration, Funding acquisition.

**Declaration of Competing Interest**

The authors declare that they have no known competing financial interests or personal relationships that could have appeared to influence the work reported in this paper.

**Data availability**

Data will be made available on request.

**Acknowledgements**

This publication is based upon work supported by the National Natural Science Foundation of China (Grant: 52350610259), Zhejiang University Global Partnership Fund, and the Zhejiang-Saudi Energy Materials International Collaboration Laboratory.

**Frequency-resolved Raman Thermometry Analysis via a Multi-layer Heat Transfer Model for Bulk and Low-dimensional Materials**

Taocheng Yu[1], Yilu Fu[1], Chenguang Fu[2], Tiejun Zhu[2], Wee-Liat Ong,[1,3]*

[1]ZJU-UIUC Institute, College of Energy Engineering, Zhejiang University, Haining, Jiaxing, Zhejiang 314400, China

[2]State Key Laboratory of Silicon and Advanced Semiconductor Materials, School of Materials Science and Engineering, Zhejiang University, 310058 Hangzhou, China.

[3]State Key Laboratory of Clean Energy Utilization, Zhejiang University, Hangzhou Zhejiang, 310027, China

**Table of Content**

## 1. Spot radius measurement

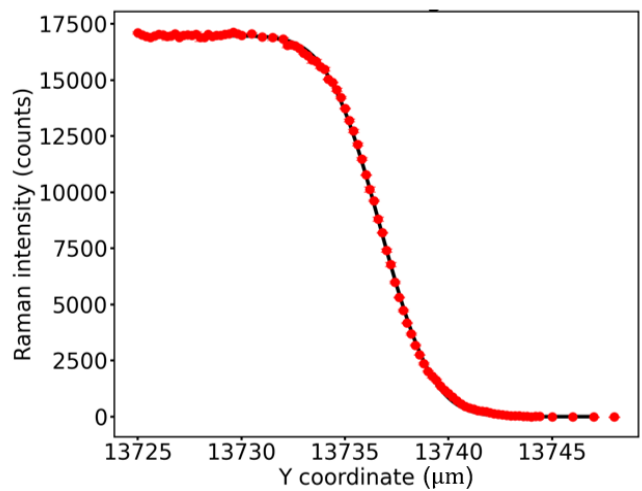

Figure S1. The measured Raman peak intensity vs position along the y-axis and fit to complementary error function (*erfc*)

We use an atomically-sharp cleaved silicon edge to translate and block the laser beam at its focal plane. The step size is 0.1μm. At each step, the Raman signal of the silicon chip is recorded three times. The average intensity of the Raman signal at each step is shown in Figure S1. The points are fitted to a complementary error function (*erfc*) to obtain a Gaussian distribution of the form, $y = \frac{1}{\sqrt{2\pi}\sigma} e^{-\frac{(x-\mu)^2}{2\sigma^2}}$. The $\sigma$ of the Gaussian distribution is 2.0±0.1μm. The $1/e^2$ spot radius is thus 4.0±0.2μm (2σ).

## 2. Laser power dependent Raman shift of bulk MoS₂ and Bi₂Se₃

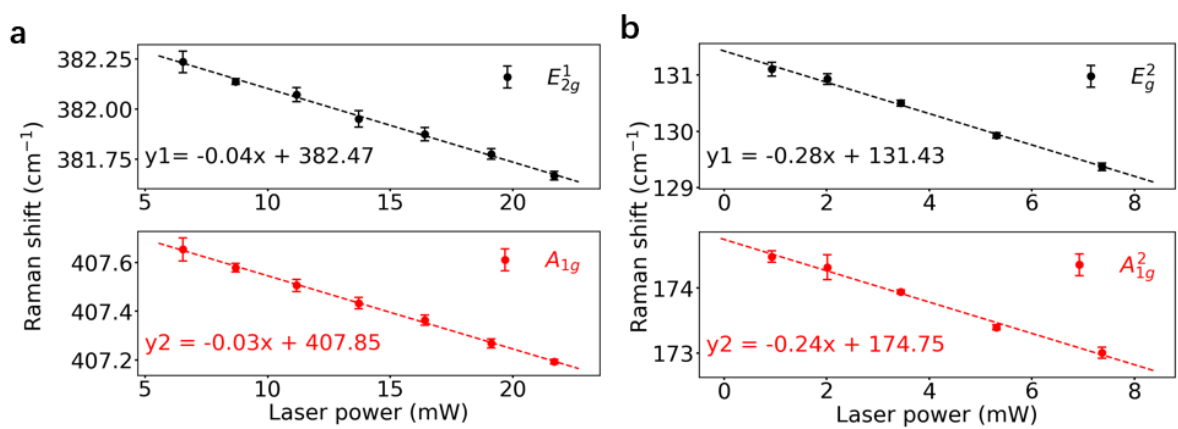

Figure S2. Raman shifts as a function of laser power of (c) bulk MoS₂ and (d) bulk Bi₂Se₃. The error bar on each point is calculated from five consecutive measurements.

## 3. Model validation

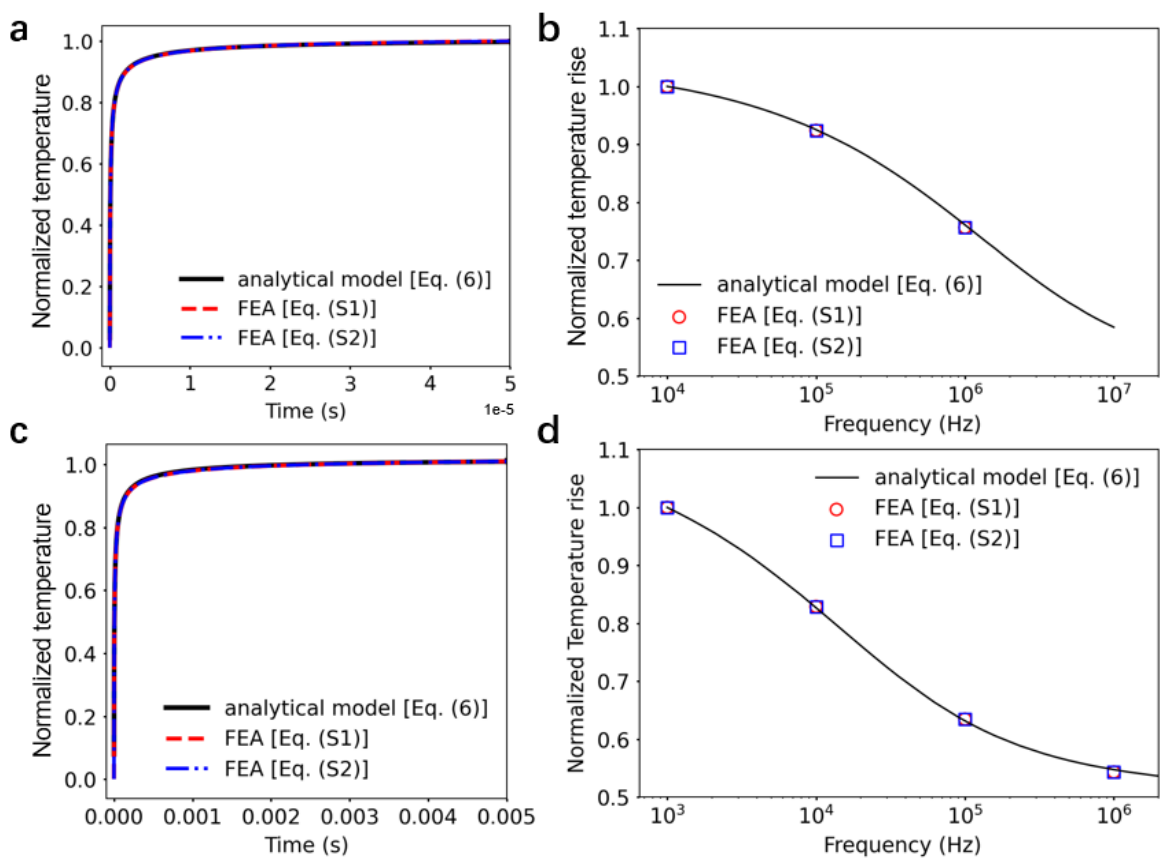

Figure S3. For bulk MoS$_2$, (a) Normalized temperature of a SW heating with a frequency of 10kHz calculated from the analytical model (black solid line) and FEA with Eq. (S1) (red dashed line) and Eq. (S2) (blue dashed line) (b) Normalized temperature of SW heating with different frequencies. For MoS$_2$/SiO$_2$, (c) Normalized temperature rise of a SW heating with a frequency of 100Hz calculated from the analytical model (black solid line) and FEA with Eq. (S1) (red dashed line) and Eq. (S2) (blue dashed line) (d) Normalized temperature rise of SW heating with different frequencies.

We validated our heat transfer model approach by comparing its result with that from the FEA. We modeled a bulk MoS$_2$ as a single-layered system and a SiO$_2$-supported MoS$_2$ thin film as a multilayered system. The finite element model is built and calculated with identical boundary conditions as the analytical model. The parameters used in both the analytical model and FEA are listed in Table S1.

The temperature probed by Raman signals is the temperature distribution weighted by the laser power intensity. The probed temperature can be written as follows,

$$H = \frac{\int_0^\infty \int_0^\infty T(r,z)I(r,z)2\pi r dr dz}{P}$$ (S1)

where $I(r,z)$ is the Gaussian beam intensity distribution along the radial (r) and exponential decay along the depth (z) direction, $T(r,z)$ is the temperature distribution, $P$ is the total absorbed laser power.

In 2D materials, the temperature gradient in the $z$ direction is negligible [1]. The probed temperature can be approximated as the surface temperature, where the temperature distribution is only weighted by the laser power intensity at the surface (z=0), as shown below,

$$H_S = \frac{\int_0^\infty T(r,0)I(r,0)2\pi r dr}{P}$$ (S2)

where subscript $s$ stands for "surface".

In our analytical model, the temperature in bulk materials is calculated from Eq. (S1) [2]. For 2D materials, the temperature is calculated from Eq. (S2) [3]. In our finite element model, temperatures from both Eq. (S1) and Eq. (S2) are recorded. Figure S3(a) and (c) shows the temperature as a function of time for bulk $MoS_2$ and $MoS_2/SiO_2$, respectively. Figure S3(b) and (d) shows the normalized temperature rise as a function of modulation frequency for bulk $MoS_2$ and $MoS_2/SiO_2$, respectively. The temperature calculated from Eq. (S1) and Eq. (S2) in FEA is represented by the red line (circle) and blue line (square), respectively. The red and blue lines (symbols) coincide with each other and both align with the black line, which represents the analytical model.

A couple of conclusions can be drawn from this. First, as our analytical model and the FEA produce almost identical transient heating process in Figures S3(a) and (c) and average temperature rise in Figures S3(b) and (d), we can use our analytical model to analyze the signal from a transient Raman thermometry measurement. Second, the difference between the temperature calculated from Eq. (S1) and Eq. (S2) is negligible. This result enables us to use Eq. (S2) to calculate the probed temperature in our analytical model.

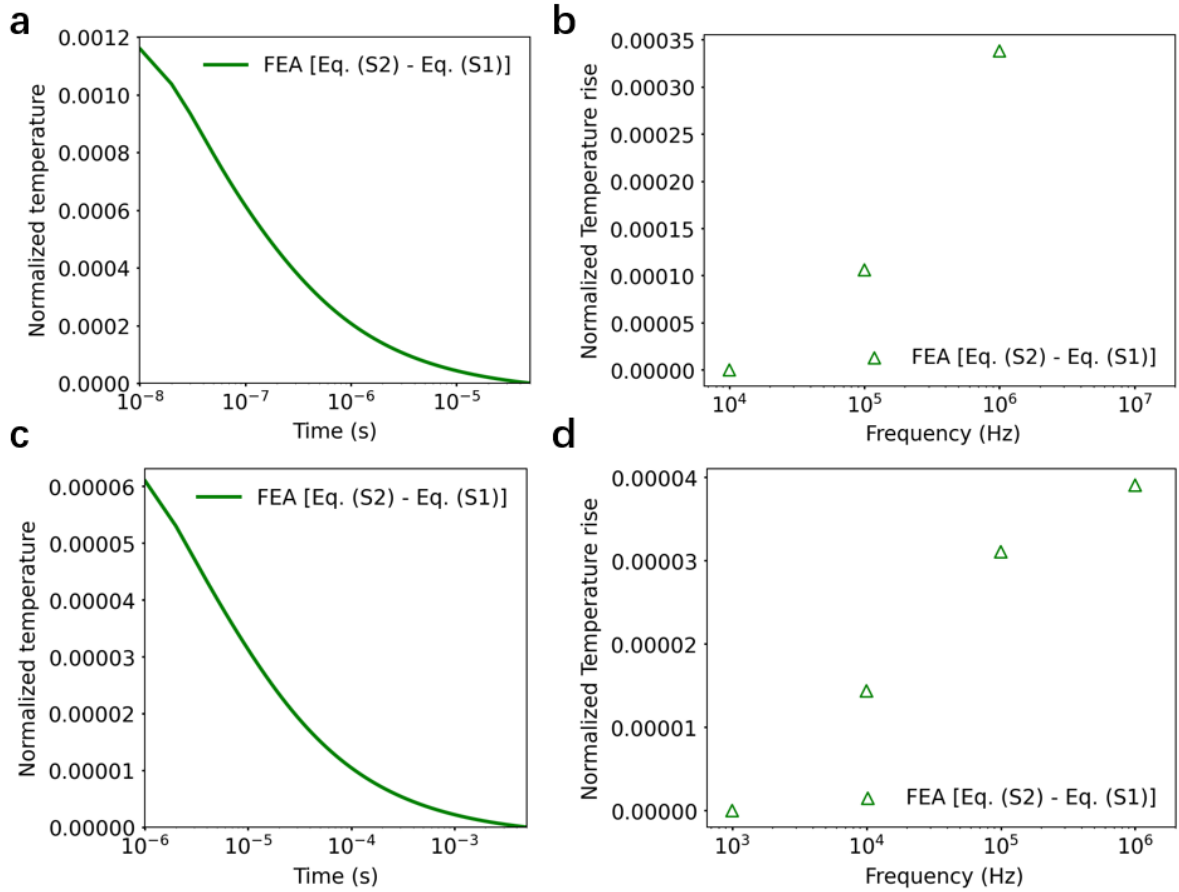

Figure S4. Difference between the FEA results calculated using Eq. (S1) and Eq. (S2). For bulk $MoS_2$, (a) the difference between the blue line and red line in Figure S1(a); (b) the difference between the blue square and red circle in Figure S1(b). For $MoS_2/SiO_2$, (c) the difference between the blue line and red line in Figure S1(c); (d) the difference between the blue square and red circle in Figure S1(d).

In Figure S4, we plotted the difference between the blue and red lines (symbols) from Figure S3. With increasing time, the difference between Eq. (S2) and Eq. (S1) decreases as heat diffuses downwards. As the frequency increases, the difference between Eq. (S2) and Eq. (S1) increases as the heating time is reduced. The magnitudes in Figure S4(b) and (d) are much smaller than one even at 1MHz, indicating that Eq. (S2) is a good approximation for Eq. (S1). Although the deviation between Eq. (S2) and Eq. (S1) is one order of magnitude larger for bulk $MoS_2$ [Figure S4(a) and (b)] than that of $MoS_2/SiO_2$ [Figure S4(c) and (d)], it is still small. This deviation comes from the upper limit of integration in the $z$ direction in Eq. (S1) which is at infinity for bulk materials, while only several nanometers for 2D materials. In this work, we use Eq. (S1) in the analytical model for bulk materials, and use Eq. (S2) as an approximation for 2D materials.

Table S1 Parameters used for the FEA and analytical model in the section of model validation

| Parameters | Values |
|---|---|
| $k_r$ of MoS$_2$ | 80 W/m/K |
| $k_z$ of MoS$_2$ | 4.75 W/m/K |
| $c$ of MoS$_2$ | 382 J/kg/K |
| $\rho$ of MoS$_2$ | 4800 kg/m$^3$ |
| $k$ of SiO$_2$ | 1.36 W/(mK) |
| $c$ of SiO$_2$ | 743 J/kg/K |
| $\rho$ of SiO$_2$ | 2200 kg/m$^3$ |
| $h$ between MoS$_2$ and SiO$_2$ | 10 MW/m$^2$/K |
| $1/e^2$ spot radius | 4 µm |
| Thickness of MoS$_2$ | 6 nm |
| Optical penetration depth of MoS$_2$ | 14 nm |

## 4. Effect of spot radius on the sensitivity

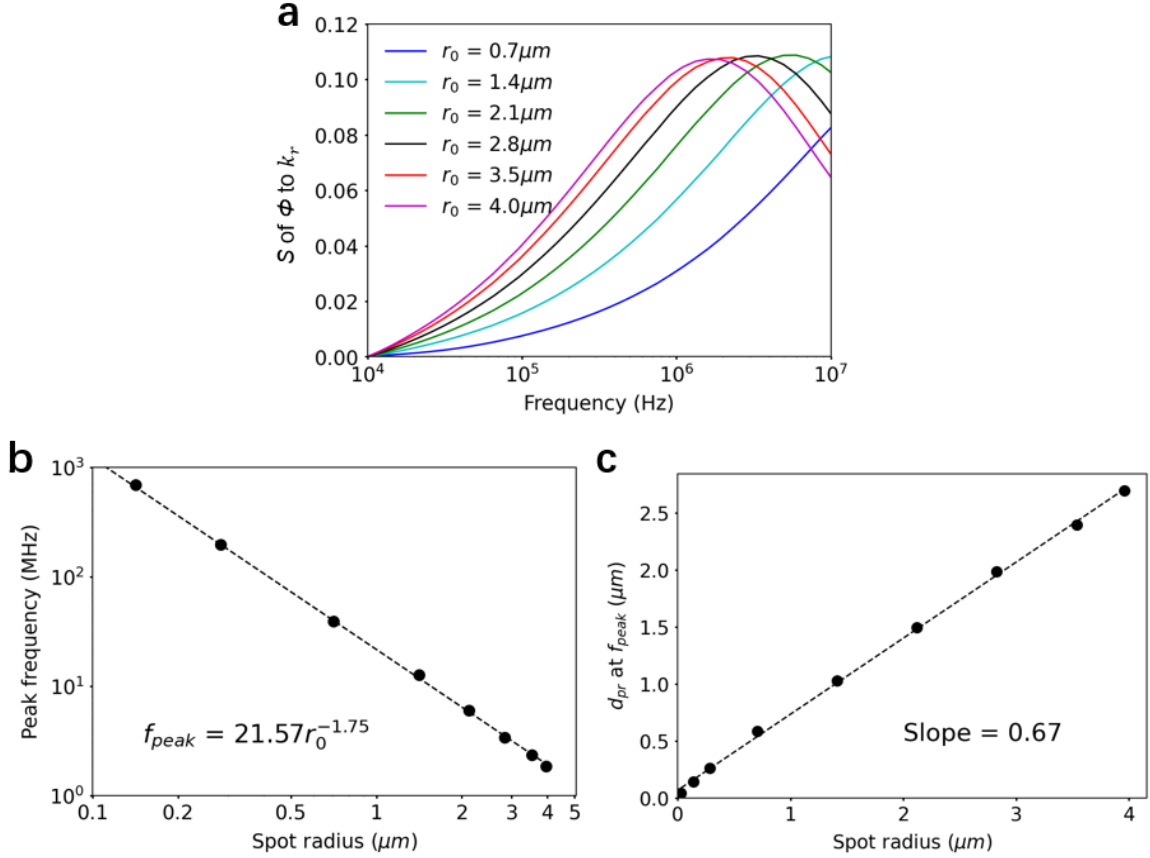

Figure S5. (a) Sensitivity of $\Phi$ to $k_r$ of MoS$_2$ with different spot sizes. (b) Peak frequency of the sensitivity curves versus spot radius. (c) In-plane heat spreading distance at peak frequency versus spot radius.

The choice of spot size is important to the measurement. We use MoS$_2$ as an example to illustrate how the spot radius $r_0$ affects the sensitivity. Figure S5(a) shows the sensitivity of $\Phi$ to $k_r$ of MoS$_2$ with different $r_0$. As $r_0$ decreases, the sensitivity curve shifts to higher frequencies. As shown in Figure S5(b), the peak frequency of the sensitivity curve ($f_{peak}$) follows the relation $f_{peak} = 21.57r_0^{-1.75}$. This relation suggests that if we use a small laser spot, we will need a high modulation frequency to include the measurement range with higher sensitivity. To get a clearer picture, the $d_{pr}$ at the $f_{peak}$ is plotted in Figure S5(c) versus its associated $r_0$, showing a linear slope of 0.67. Thus, for a given $r_0$, the $f$ range should include the frequency that produces $d_{pr}/r_0$=0.67. So, it is important to select an appropriate spot size according to the modulated frequency.

## 5. Optical-acoustic phonon nonequilibrium

The temperature $T$ in Eq. (1) is the temperature of optical phonons (OP) while the major heat carrier is acoustic phonons (AP). We directly compare the $\Phi$ to acoustic phonon temperature rise ($\Delta T_{AP}$) assuming OP and AP are at equilibrium ($\Delta T_{AP} = \Delta T_{OP}$). Since OP transfer energy to AP under laser irradiation, there will be a temperature difference between OP and AP ($\Delta T_{OA}$). The contribution of $\Delta T_{OA}$ to $\Delta T_{OP}$ ($\Delta T_{OA}\%$) decreases as the laser spot size increases [4]. We calculate $\Delta T_{OA}\%$ using the method developed by Hunter et al. [5] using the Raman shift coefficient, $\psi$. This coefficient is obtained by linear fitting the relation between Raman shift and laser power, measured with 20X and 50X objective lenses. Since $\psi$ represents the OP temperature rise under unit laser power, it is proportional to the total resistance and can be written as $\psi = \frac{A}{r_0 + \Delta r} + \frac{B}{(r_0 + \Delta r)^2} + C/r_0^2$ [5]. Here, $A$, $B$ and $C$ are constants, and $\Delta r$ accounts for spot radius enlargement due to hot carrier diffusion. The term $C/r_0^2$ represents the contribution of OA to $\psi$. We normalize the $\psi$ of different $r_0$ with $\psi_{20X}$, and get $\Omega = \psi/\psi_{20X}$.

A new parameter $\Omega r_0^2$ is defined. As stated in [5], $\Omega r_0^2$ can be approximated as linear to $r_0$ when $\Delta r$ is small. A constant $C'$ can be determined from linearly fitting the $\Omega r_0^2 - r_0$ relationship and get the intercept. $C'$ is determined to be 0.4872 and 0.8066 for MoSe$_2$/SiO$_2$ and MoSe$_2$/SiO$_2$, respectively. The percentage of $\Delta T_{OA}$ to $\Delta T_{OP}$ is $C'/r_0^2$, which is 3.1% and 5.1% for MoSe$_2$/SiO$_2$ and MoSe$_2$/SiO$_2$, respectively.

The higher $\Delta T_{OA}\%$ of MoSe$_2$/SiO$_2$ than that of MoS$_2$/SiO$_2$ implies a weaker optical-acoustic phonon coupling for the former, consistent with a prior study [6]. The change in Raman shift in Eq. (1) is multiplied by (1-$\Delta T_{OA}\%$) to get the normalized acoustic phonon temperature rise ($\Phi_{AP}$). The $\Phi_{AP}$ is then fitted to the heat transfer model to get $h$.

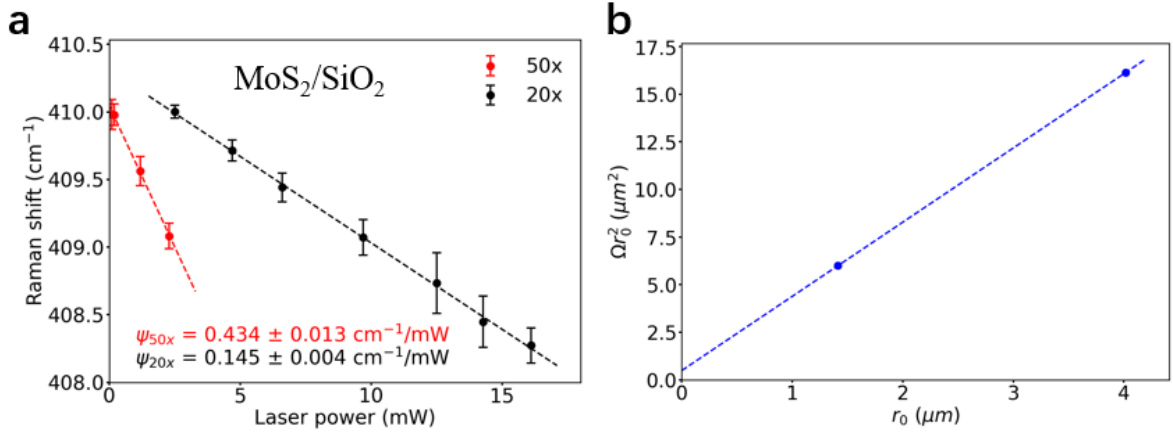

Figure S6. (a) Linear relation between the Raman shift and laser power of MoS₂/SiO₂. (b) linear fitting of $\Omega r_0^2 - r_0$ relation of MoS₂/SiO₂

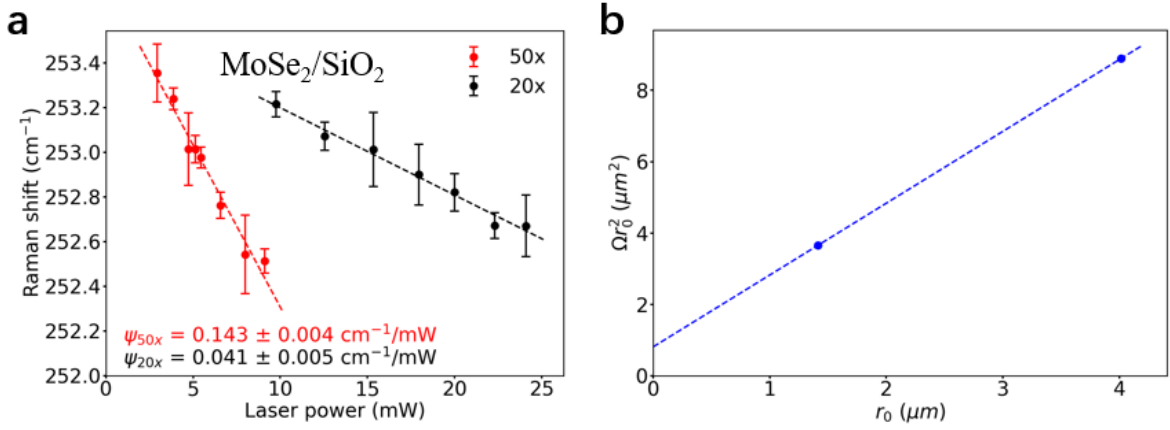

Figure S7. (a) Linear relation between the Raman shift and laser power of MoSe₂/SiO₂. (b) Linear fitting of $\Omega r_0^2 - r_0$ relation of MoSe₂/SiO₂

## 5. Calculations of the diffuse mismatch model (DMM)

The diffuse mismatch model (DMM) is used to calculate the interfacial thermal conductance ($h$). The thermal conductance across an interface between A and B is defined as the ratio of the heat current density and the temperature differential. Following the expression formulated by Reddy et al. [7], $h$ can be calculated by Eq. (S3)

$$ h = \frac{1}{2(2\pi)^3} \sum_i \int \frac{1}{k_B T^2} \alpha_{A \to B}(k,i) \times (\hbar\omega(k,i))^2 |V(k,i) \cdot n| \frac{\exp\left(\frac{\hbar\omega(k,i)}{k_B T}\right)}{[\exp\left(\frac{\hbar\omega(k,i)}{k_B T}\right)-1]^2} \, dk, \qquad (S3) $$

where $\alpha_{A \to B}(k,i)$ is the transmission probability of A to B, $\omega(k,i)$ and $V(k,i)$ are the phonon frequency and group velocity corresponding to wave vector $k$ and phonon mode $i$ in medium A, and $n$ is the unit vector normal to the interface from A into B. The transmission probability is calculated from the group velocity as follows,

$$\alpha_{A \to B}\left(\omega^{'}\right) = \frac{\Delta K_B[\sum_{j,k}|V(k,j)\cdot n|]\delta_{\omega(k,j),\omega^{'}}}{\Delta K_A[\sum_{i,k}|V(k,i)\cdot n|]\delta_{\omega(k,i),\omega^{'}} + \Delta K_B[\sum_{j,k}|V(k,j)\cdot n|]\delta_{\omega(k,j),\omega^{'}}}, \tag{S4}$$

where $\Delta K_A$ and $\Delta K_B$ are the volumes of discretized cells in the Brillouin zone of A and B, respectively, and $\delta_{\omega(k,j),\omega^{'}}$ is the Kronecker delta function which is equal to unity when $\omega(k,j) = \omega^{'}$ and zero otherwise. The summation is carried out over the first Brillouin zone for all modes. Eqs. (S3) and (S4) can be used when the full phonon dispersion is known.

The phonon dispersions and phonon density of states are obtained using the first-principles calculations. We employ the QUANTUM ESPRESSO package with PBE optimized norm-conserving Vanderbilt pseudopotentials. The atomic positions are relaxed using an electronic wave-vector grid of 5 × 5 × 2 (4 × 4 × 4) for $MoS_2$/$MoSe_2$ ($SiO_2$) to ensure the residual forces on each atom are less than $10^{-5}$ Ry/Å. The plane wave cutoff energy is set to 60 Ry in all calculations. The second order interatomic force constants (IFCs) are obtained using density functional perturbation theory (DFPT). The calculated phonon dispersions are shown in Figure S8. To do the summation in Eq. (S3) and (S4), a 6 × 6 × 10 grid is used to sample the $k$ points in the irreducible Brillouin zone, which has a tested to be dense enough to get the converged result.

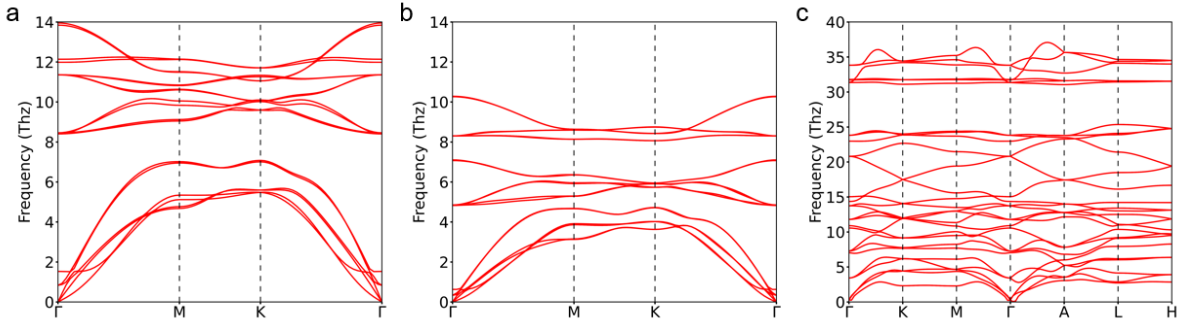

Figure S8. Phonon dispersion of (a) $MoS_2$, (b) $MoSe_2$, and (c) $SiO_2$.

## 6. Sensitivity curve for a 100 nm thick MoS₂/SiO₂ system

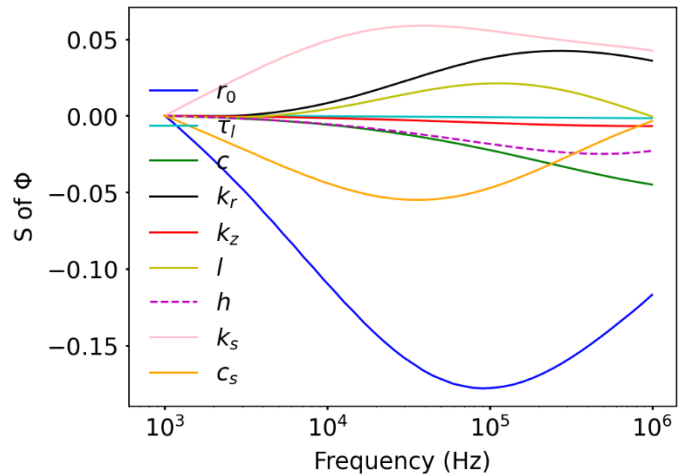

Figure S9. Sensitivity curve for a 100 nm thick MoS₂/SiO₂ system.